\documentclass{ieeeaccess}
\usepackage{amsmath,amssymb,amsfonts}
\usepackage{graphicx}
\usepackage{textcomp}

\graphicspath{ {./figures/} }
\usepackage[pagebackref,breaklinks,bookmarks=false,colorlinks,linkcolor=black,anchorcolor=black,citecolor=black,filecolor=black,menucolor=black,runcolor=black,urlcolor=black]{hyperref}
\usepackage{amsthm}
\usepackage[numbers]{natbib}
\theoremstyle{definition}
\newtheorem{definition}{Definition}[]
\newtheorem{theorem}{Theorem}
\usepackage[ruled,vlined, linesnumbered]{algorithm2e}
\usepackage{booktabs}

\newcommand{\orcidicon}[1]{\href{https://orcid.org/#1}{\includegraphics[height=\fontcharht\font`\B]{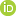}}}

\begin{document}
\history{Date of publication xxxx 00, 0000, date of current version November 28, 2021.}
\history{Received November 28, 2021, accepted December 27, 2021, date of publication xxx, date of current version December 27, 2021.}
\doi{10.1109/ACCESS.2021.3139510}

\title{Reinforcement Learning for Systematic FX Trading}
\author{\uppercase{Gabriel Borrageiro}\orcidicon{0000-0002-0063-7103}, 
\uppercase{Nick Firoozye\orcidicon{0000-0002-6460-0406} and Paolo Barucca\orcidicon{0000-0003-4588-667X}}}
\address{Department of Computer Science, University College London, Gower Street, London, WC1E 6BT}

\markboth
{Borrageiro et al.: Reinforcement Learning for Systematic FX Trading}
{Borrageiro et al.: Reinforcement Learning for Systematic FX Trading}

\corresp{Corresponding author: Gabriel Borrageiro (e-mail: gabriel.borrageiro.20@ucl.ac.uk).}

\begin{abstract}
We explore online inductive transfer learning, with a feature representation transfer from a radial basis function network formed of Gaussian mixture model hidden processing units to a direct, recurrent reinforcement learning agent. This agent is put to work in an experiment, trading the major spot market currency pairs, where we accurately account for transaction and funding costs. These sources of profit and loss, including the price trends that occur in the currency markets, are made available to the agent via a quadratic utility, who learns to target a position directly. We improve upon earlier work by targeting a risk position in an online transfer learning context. Our agent achieves an annualised portfolio information ratio of 0.52 with a compound return of 9.3\%, net of execution and funding cost, over a 7-year test set; this is despite forcing the model to trade at the close of the trading day at 5 pm EST when trading costs are statistically the most expensive.
\end{abstract}


\titlepgskip=-15pt

\maketitle
\section{Introduction}
Forecasters of financial time series commonly make use of supervised learning. For example, \cite{TsayRueyS.2019Ntsa} apply both parametric approaches such as nonlinear state-space models and non-parametric approaches such as local learning to nonlinear time series analysis. \cite{Bengio1997} applies learning algorithms to decision making with financial time series. He notes that the traditional approach in this domain is to train a model using a prediction criterion, such as minimising mean-square prediction error or maximising the likelihood of a conditional model of the dependent variable. He finds that with noisy time series, better results are obtained when the model is trained directly to maximise the financial criterion of interest, here gains and losses (including those due to transactions) incurred during trading.

In this spirit, we extend the earlier work of \cite{MoodyWu1997} and \cite{Gold2003}, where direct, recurrent reinforcement learning agents are put to work in financial trading strategies. Rather than optimising for an intermediate performance measure such as maximal forecast accuracy or minimal forecast error, which is still the traditional approach in this domain, we maximise a more direct performance measure such as quadratic economic utility. An advantage of the approach is that we can use the risk-adjusted returns of the trading strategy, execution cost and funding cost to influence the learning of the model and update model parameters accordingly. 

Whereas the focus of \cite{MoodyWu1997} was on the use of the differential Sharpe ratio as a performance measure, we adopt the quadratic utility of \cite{Sharpe2007}. This utility ameliorates the undesirable property of the Sharpe ratio in that it penalises a model that produces returns larger than $\frac{\mathbb{E}[r^2_t]}{\mathbb{E}[r_t]}$, that is, the ratio of the expectation of squared returns to the expectation of returns \citep{Moody1998PerformanceFA}. For this reason, along with the use of relatively weak features and shared backtest hyper-parameters, \cite{Gold2003} obtained mixed results when experimenting with cash currency pairs. In contrast, our experiment with the major cash currency pairs sees our recurrent reinforcement learning trader achieve an annualised portfolio information ratio of 0.52 with a compound return of 9.3\%, net of execution and funding cost, over a seven-year test set. This return is achieved despite forcing the model to trade at the close of the trading day at 5 pm EST when trading costs are statistically the most expensive. 

Aside from the different utility functions, we put these improved experiment results down to a combination of several factors. Firstly, we use more powerful feature engineering in the shape of radial basis function networks. The hidden processing units of these networks have means, covariances and structures that are determined by an unsupervised learning procedure for finite Gaussian mixture models \citep{FigueiredoJain}. The approach is a form of continual learning, explicitly inductive, feature representation transfer learning \citep{yang_transfer_learning_2020}, where the knowledge of the mixture model is transferred to upstream models. Secondly, when optimising our utility function with respect to the recurrent reinforcement learner's parameters, we do so sequentially online during the test set, using an extended Kalman filter optimisation procedure \citep{Haykin2001}. The earlier work uses less powerful offline batch gradient ascent methods. These methods cope less well with non-stationary financial time series. 

\cite{Merton1976} modelled the dynamics of financial assets as a jump-diffusion process, which is commonly used in financial econometrics. The jump-diffusion process implies that financial time series should observe small changes over time, so-called continuous changes, as well as occasional jumps. A sensible approach for coping with nonstationarity is to allow models to learn continuously.

We finish this section with a description of the layout of this paper. Section \ref{sec:prelim} provides preliminary introductions to transfer learning and reinforcement learning via policy gradients and ends with an overview of trading in the foreign exchange market. Section \ref{sec:methods} introduces the experiment methods of this paper, including the targeting of financial risk positions with direct recurrent reinforcement and feature representation transfer via radial basis function networks. The section ends with a description of the baseline models used to compare the results of the marquis model. The marquis model is a feature representation transfer from a radial basis function network to a direct recurrent reinforcement learning agent and is shown visually in figure \ref{fig:rbf_net}. 

Section \ref{sec:experiment_design} details the design of the experiment that we conduct on daily sampled foreign exchange pairs. The data is obtained from Refinitiv. We evaluate performance using the annualised information ratio, which is computed on daily returns that are net of transaction and funding costs. The section completes a brief description of the hyper-parameters set for the various models. The experiment results are described in section \ref{sec:results} and are discussed in section \ref{sec:discussion}. Concluding remarks are given in section \ref{sec:conclusion}.

\section{Preliminaries} \label{sec:prelim}
This section introduces the policy gradient form of reinforcement learning and how it has been put to work empirically in quantitative finance, particularly with automated trading strategies. Finally, we finish the section with a short review of more recent work.

\subsection{Transfer Learning} \label{sec:transfer_learning}
Transfer learning refers to the machine learning paradigm in which an algorithm extracts knowledge from one or more application scenarios to help boost the learning performance in a target scenario \cite{yang_transfer_learning_2020}. Typically, traditional machine learning requires significant amounts of training data. Transfer learning copes better with data sparsity by looking at related learning domains where data is sufficient. Even in a big data scenario such as with streaming high-frequency data, transfer learning benefits by learning the adaptive statistical relationship of the predictors and the response. An increasing number of papers focus on online transfer learning \cite{Zhao_OTL_2014,Bruno_OTL_2019,Wang_OTL_2020}. Following Pan and Yang \cite{Pan2010ASO}, we define transfer learning as: 
\begin{definition}[transfer learning]
Given a source domain $\mathcal{D}_S$ and learning task $\mathcal{T}_S$, a target domain $\mathcal{D}_T$ and learning task $\mathcal{T}_T$, transfer learning aims to help improve the learning of the target predictive function $f_T(.)$ in $\mathcal{D}_T$ using the knowledge in $\mathcal{D}_S$ and $\mathcal{T}_S$, where $\mathcal{D}_S \neq \mathcal{D}_T$, or $\mathcal{T}_S \neq \mathcal{T}_T$.
\end{definition}

In the context of this paper, the source domain $\mathcal{D}_S$ represents the feature space, which consists of the daily returns of the 36 currency pairs that are used in our experiment. The source learning task $\mathcal{T}_S$ is the unsupervised compression of this feature space into a clustered form that learns its intrinsic nature. The clusters are formed via Gaussian mixture models, and we transfer their output via radial basis function networks to currency pairs that we wish to trade in the target domain $\mathcal{D}_T$. The target learning task $\mathcal{T}_T$ is to take financial risk positions in these currency pairs for economic utility maximisation via direct recurrent reinforcement learning.

\subsection{Policy Gradient Reinforcement Learning} \label{sec:policy_gradients}
Williams \cite{Williams1992} was one of the first to introduce policy gradient methods in a reinforcement learning context. Whereas most reinforcement learning algorithms focus on action-value estimation, learning the value of actions and selecting them based on their estimated values, policy gradient methods learn a parameterised policy that can select actions without using a value function. Williams also introduced his \textit{reinforce} algorithm
\[
\Delta \boldsymbol{\theta}_{ij} = \eta_{ij}(r - b_{ij})\ln (\partial{\pi_i} / \partial{\boldsymbol{\theta}}_{ij}),
\]
where $\boldsymbol{\theta}_{ij}$ is the model weight going from the $j'th$ input to the $i'th$ output, and $\boldsymbol{\theta}_i$ is the weight vector for the $i'th$ hidden processing unit of a network of such units, whose goal it is to adapt in such a way as to maximise the scalar reward $r$. For the moment, we exclude the dependence on the time of the weight update to make the notation clearer. Furthermore, $\eta_{ij}$ is a learning rate, typically applied with gradient ascent, $b_{ij}$ is a reinforcement baseline, conditionally independent of the model outputs $y_i$, given the network parameters $\boldsymbol{\theta}$ and inputs $\mathbf{x}_i$. $\ln (\partial{\pi_i} / \partial{\boldsymbol{\theta}}_{ij})$ is known as the characteristic eligibility of $\boldsymbol{\theta}_{ij}$, where $\pi_i(y_i = c, \boldsymbol{\theta}_i, \mathbf{x}_i)$ , is a probability mass function determining the value of $y_i$ as a function of the parameters of the unit and its input. Baseline subtraction $r - b_{ij}$ plays a vital role in reducing the variance of gradient estimators. Sugiyama \cite{Sugiyama2015} shows that the optimal baseline is given as
\[
b^* = \frac{\mathbb{E}_{p(r|\boldsymbol{\theta})} \big [ r_t \| \sum_{t=1}^T \nabla \ln{\pi(a_t|s_t, \boldsymbol{\theta}) \|^2} \big ]}{\mathbb{E}_{p(r|\boldsymbol{\theta})} \big [\| \sum_{t=1}^T \nabla \ln{\pi(a_t|s_t, \boldsymbol{\theta}) \|^2} \big ]},
\]
where the policy function $\pi(a_t|s_t, \boldsymbol{\theta})$ denotes the probability of taking action $a_t$ at time t given state $s_t$, parameterised by $\boldsymbol{\theta}$. The expectation $\mathbb{E}_{p(r|\boldsymbol{\theta})}$, is distributed over the probability of rewards given the model parameterisation.

The main result of Williams' paper is 
\begin{theorem}
For any \textit{reinforce} algorithm, the inner product of $\mathbb{E}[\Delta \boldsymbol{\theta} | \boldsymbol{\theta}]$ and $\nabla \mathbb{E}[r | \boldsymbol{\theta}]$ is non-negative, and if $\eta_{ij} > 0$, then this inner product is zero if and only if $\nabla \mathbb{E}[r | \boldsymbol{\theta}] = 0$. If $\eta_{ij}$ is independent of $i$ and $j$, then $\mathbb{E}[\Delta \boldsymbol{\theta} | \boldsymbol{\theta}] = \eta \nabla \mathbb{E}[r | \boldsymbol{\theta}]$.
\end{theorem}
This result relates $\nabla \mathbb{E}[r | \boldsymbol{\theta}]$, the gradient in weight space of the performance measure $\mathbb{E}[r| \boldsymbol{\theta}]$, to $\mathbb{E}[\Delta \boldsymbol{\theta} | \boldsymbol{\theta}]$, the average update vector in weight space. Thus for any \textit{reinforce} algorithm, the average update vector in weight space lies in a direction for which this performance measure is increasing, and the quantity $(r - b_{ij})\ln (\partial{\pi_i} / \partial{\boldsymbol{\theta}}_{ij})$ represents an unbiased estimate of $\partial{\mathbb{E}[r|\boldsymbol{\theta}}] / \partial{\boldsymbol{\theta}_{ij}}$.

Sutton and Barto \cite{SuttonBarto2018} demonstrate an \textit{actor-critic} version of a policy gradient model, where the actor references the learned policy and the critic refers to the learned value function, usually a state-value function. Denote the scalar performance measure as $J(\boldsymbol{\theta})$; the gradient ascent update takes the form
\[
\boldsymbol{\theta}_{t+1} = \boldsymbol{\theta}_t + \eta \nabla J(\boldsymbol{\theta}).
\]
With the one-step actor-critic policy gradient algorithm, one inserts a differentiable policy parameterisation $\pi(a|s,\boldsymbol{\theta})$, a differentiable state-value function parameterisation $\hat{v}(s, \mathbf{w})$ and then one draws an action
\[
a_t \sim \pi(.|s_t, \boldsymbol{\theta}),
\]
taking action $a_t$ and observing a transition to state $s_{t+1}$ with reward $r_{t+1}$. Define
\[
\delta_t = r_{t+1} + \gamma \hat{v}(s_{t+1}, \mathbf{w_t}) - \hat{v}(s_{t}, \mathbf{w_t}),
\]
where $0 \ll \gamma \leq 1$ is discount factor. The critic's weight vector is updated as follows
\[
\mathbf{w}_t = \mathbf{w}_{t-1} + \eta_{\mathbf{w}} \delta_t \nabla_{\mathbf{w}} \hat{v}(s_{t}, \mathbf{w_t}),
\]
and finally, the actor's weight vector is updated as
\[
\boldsymbol{\theta}_t = \boldsymbol{\theta}_{t-1} + \eta_{\boldsymbol{\theta}} \delta_t \nabla \ln \pi(a_t|s_t, \boldsymbol{\theta}).
\]
The actor-critic architecture uses temporal-difference learning combined with trial-and-error learning to improve the learned policy sequentially.

\subsubsection{Policy Gradient Methods in Financial Trading}
Moody et al. \cite{Moody1998PerformanceFA} propose to train trading systems and portfolios by optimising objective functions that directly measure trading and investment performance. Rather than basing a trading system on forecasts or training via a supervised learning algorithm using labelled trading data, they train their systems using a direct, recurrent reinforcement learning algorithm, an example of the policy gradient method. The \textit{direct} part refers to the fact that the model tries to target a position directly, and the model's weights are adapted such that the performance measure is maximised. The performance function that they primarily consider is the differential Sharpe ratio. Denote the annualised Sharpe ratio \cite{Sharpe1966} as
\[
 sr_k = 252^{0.5} \times \frac{r_k - r_f}{s_k},
\]
where $r_k$ is the return of the $k'th$ strategy, with standard deviation $s_k$ and $r_f$ is the risk-free rate. For ease of explanation, we now remove the strategy index $k$ and replace it with a time index $t$. The differential Sharpe ratio is defined as
\begin{equation}
 \frac{dsr_t}{d\tau} = \frac{b_{t-1} \Delta a_t - 0.5 a_{t-1} \Delta b_t}
 {(a_{t-1} - a_{t-1}^2)^{3/2}},
\end{equation}
where the quantities $a_t$ and $b_t$ are exponentially weighted estimates of the first and second moments of the reward $r_t$
\[
 \begin{aligned}
 a_t &= a_{t-1} + \tau \Delta a_t = a_{t-1} + \tau (r_t - a_{t-1}) \\
 b_t &= b_{t-1} + \tau \Delta b_t = b_{t-1} + \tau (r_t^2 - b_{t-1}).
 \end{aligned}
\]
The exponential decay constant is $\tau \in (0, 1]$. They consider a batch gradient ascent update for model parameters $\boldsymbol{\theta}$
\[
\Delta{\boldsymbol{\theta}_T = \eta \frac{dsr_T}{d\boldsymbol{\theta}}},
\]
where
\[
\begin{split}
 \frac{dsr_T}{d\boldsymbol{\theta}} & = \sum_{t=1}^T \frac{dsr_T}{dr_t} \frac{dr_t}{d\boldsymbol{\theta}} \\
 & = \sum_{t=1}^T \Bigg\{ \frac{b_T - a_T r_t}{(b_T - a_T^2)^{3/2}} \Bigg\} \Bigg\{ \frac{dr_t}{df_t} \frac{df_t}{d\boldsymbol{\theta}} + \frac{dr_t}{df_{t-1}} \frac{df_{t-1}}{d\boldsymbol{\theta}} \Bigg\}.
\end{split}
\]
The reward 
\[
r_t = \Delta p_t f_{t-1} - \delta_t|\Delta f_t|
\]
depends on the change in reference price $p_t$ from which a gross profit and loss are computed, transaction cost $\delta_t$ and a differentiable position function of the model inputs and parameters $f_t \triangleq f(\mathbf{x}_t, \boldsymbol{\theta}_t)$ which is in the range $-1 \leq f_t \leq 1$. 

Trading and portfolio management systems require prior decisions as input to properly consider the effect of transaction costs, market impact, and taxes. This temporal dependence on the system state requires reinforcement versions of standard recurrent learning algorithms. Moody et al. \cite{Moody1998PerformanceFA} present empirical results in controlled experiments that demonstrate the efficacy of some of their methods for optimising trading systems and portfolios. For a long/short trader, they find that maximising the differential Sharpe ratio yields more consistent results than maximising profits. Both methods outperform a trading system based on forecasts that minimise mean-square error. They find that portfolio trading agents trained to maximise the differential Sharpe ratio achieve better risk-adjusted returns than those trained to maximise profit. However, an undesirable property of the Sharpe ratio is that it penalises a model that produces returns larger than $\frac{\mathbb{E}[r^2]}{\mathbb{E}[r]} \approx \frac{b_t}{a_t}$, that is, the ratio of the expectation of squared returns to the expectation of returns, which is counter-intuitive to investors' notion of risk and reward.

Gold \cite{Gold2003} extends Moody et al.'s \cite{Moody1998PerformanceFA} work and investigates high-frequency currency trading with neural networks trained via recurrent reinforcement learning. He compares the performance of linear networks with neural networks containing a single hidden layer and examines the impact of shared system hyper-parameters on performance. In general, he concludes that the trading systems may be effective but that the performance varies widely for different currency markets, and simple statistics of the markets cannot explain this variability. 

He also finds that the linear recurrent reinforcement learners outperform the neural recurrent reinforcement learners in this application. Here, we suspect that the choice of inputs (past returns of the target) results in features with weak predictive power. As a result, the neural reinforcement learner struggles to make meaningful forecasts. In comparison, the linear recurrent reinforcement learner does better coping with both noisy inputs and outputs, generating biased yet stable predictions. Gold also used shared hyper-parameters. Many of the currency pairs behave differently in terms of their price action. For example, US dollar crosses are usually momentum-driven. Cross-currencies, such as the Australian dollar versus the New Zealand dollar, tend to be mean-reverting in nature. Therefore, sharing hyper-parameters probably negatively impacts the ex-post performance here.

\subsubsection{More Recent Work}
In terms of more recent work involving policy gradient methods in finance, Tamar et al. \cite{TamarAviv2017SDMW} discuss risk-sensitive policy gradient methods that augment the standard expected cost minimisation problem with a measure of variability in cost. They consider static and time-consistent dynamic risk measures that combine a standard sampling approach with convex programming. Their approach is actor-critic for dynamic risk measures and involves explicit approximation of value functions.

Luo et al. \cite{LuoSuyuan2019AnCb} build a novel reinforcement learning framework trader. They adopt an actor-critic algorithm called deep deterministic policy gradient to find the optimal policy. Their proposed algorithm uses convolutional neural networks and outperforms some baseline methods when experimenting with stock index futures. They also discuss the generalisation and implications of the proposed method for finance.

Zhang et al. \cite{Zhang2019DeepRL} use deep reinforcement learning algorithms such as deep q-learning networks \cite{mnih2013playing}, neural policy gradients \cite{Silver2016} and advantage actor-critic \cite{mnih2016asynchronous} to design trading strategies for continuous futures contracts. They use long short-term memory neural networks \cite{hochreiter1997long} to train both the actor and critic networks. Both discrete and continuous action spaces are considered, and volatility scaling is incorporated to create reward functions that scale trade positions based on market volatility. They show that their method outperforms various baseline models, delivering positive profits despite high transaction costs. Their experiments show that the proposed algorithms can follow prominent market trends without changing positions and scale down or hold through consolidation periods.

Azhikodan et al. \cite{Azhikodan2019} propose automated trading systems that use deep reinforcement learning, specifically a deep deterministic policy gradient-based neural network model that trades stocks to maximise the gain in asset value. They determine the need for an additional system for trend-following to work alongside the reinforcement learning algorithm. Thus they implement a sentiment analysis model using a recurrent convolutional neural network to predict the stock trend from financial news. 

Ye et al. \cite{YeZekun2020OTEB} address an optimal trade execution problem that involves limit order books. Here, the model must learn how best to execute a given block of shares at minimal cost or maximal return. To this end, they propose a deep reinforcement learning-based solution that uses a deterministic policy gradient framework. Experiments on three real market datasets show that the proposed approach significantly outperforms other methods such as a submit and leave policy, a q-learning algorithm \cite{Watkins89} and a hybrid method that combines the Almgren-Chriss model \cite{almgren2001optimal} with reinforcement learning.

Aboussalah and Lee \cite{ABOUSSALAH2020112891} explore policy gradient techniques for continuous action and multi-dimensional state spaces, applying a stacked deep dynamic recurrent reinforcement learning architecture to construct an optimal real-time portfolio. The algorithm adopts the Sharpe ratio as a utility function to learn the market conditions and rebalance the portfolio accordingly.

Betancourt and Chen \cite{BETANCOURT2021114002} propose a novel portfolio management method using deep reinforcement learning on markets with a dynamic number of assets. Their model endeavours to learn the optimal inventory to hold whilst minimising transaction costs.

Lei et al. \cite{LEI2020112872} acknowledge that algorithmic trading is an ongoing decision making problem, where the environment requires agents to learn feature representation from highly non-stationary and noisy financial time series, and decision making requires that agents explore the environment and simultaneously make correct decisions in an online manner without any supervised information. Instead, they propose to tackle both problems via a time-driven feature-aware deep reinforcement learning model to improve the financial signal representation learning and decision making. 

\subsection{Foreign Exchange Trading} \label{sec:fx_trading}
This section describes the foreign exchange market and the mechanics of the foreign exchange derivatives, which are central to the experimentation that we conduct in section \ref{sec:experiment_design}. The global foreign exchange market sees transactions above 6 trillion US dollars traded daily. Figure \ref{fig:bis2019} shows this breakdown by instrument type and is extracted from the Bank of International Settlements Triennial Central Bank Survey, 2019.

\Figure[!t]()[width=0.99\columnwidth]{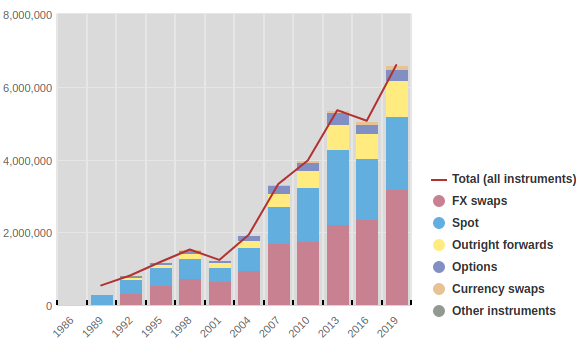}
 {Average daily global foreign exchange market turnover in millions of US dollars, source: Bank of International Settlements.\label{fig:bis2019}}

FX transactions implicitly involve two currencies: the dominant or base currency is quoted conventionally on the left-hand side and the secondary or counter currency on the right-hand side. If foreign exchange positions are held overnight, the trader will earn the interest rate of the currency bought and pay the interest rate of the currency sold. The interest rates for specific maturities are determined in the inter-bank currency market and are heavily influenced by the base rates typically set by central banks. Foreign exchange trades settle two business days after the trade date by market convention unless otherwise specified. 

Clients fund their positions by rolling them forward via tomorrow/next (tomnext) swaps. Tomnext is a short-term foreign exchange transaction where a currency pair is simultaneously bought and sold over two business days: tomorrow (in one business day) and the following day (two business days from today). The tomnext transaction allows traders to maintain their position without being forced to take physical delivery and is the convention applied by prime brokers to their clients on the inter-bank foreign exchange market. In order to determine this funding cost, one needs to compute the forward rates (prices). Forwards are agreements between two counterparties to exchange currencies at a predetermined rate on some future date. 

Forward rates are calculated by adding forward points to a spot rate. These points reflect the interest rate differential between the two currencies being traded and the maturity of the trade. Forward points do not represent an expectation of the direction of a currency but rather the interest rate differential. Let $bid_t^{spot}$ denote the spot/cash currency pair rate at which price takers can sell at time $t$. Similarly, let $ask_t^{spot}$ denote the spot/cash currency pair rate at which price takers can buy at time $t$. The spot mid-rate is 
\begin{equation}\label{eq:midprice}
 mid_t^{spot} = 0.5 \times (bid_t^{spot} + ask_t^{spot}).
\end{equation}

Forward points are computed as follows
\[
mid_t^{fpts} = mid_t^{spot}(e_2 - e_1)\frac{T}{360 \phi},
\]
where $e_2$ is the secondary interest rate, $e_1$ is the dominant interest rate, $T$ is the number of days till maturity, and $\phi$ is the tick size or pip value for the associated currency pair. Example forward points for GBPUSD are shown in figure \ref{fig:gbpusd_fwd_rates}. \textit{GBP=} is the Refinitiv information code (ric) for cash GBPUSD and \textit{GBPTND=} is the ric for tomnext GBPUSD forward points. Note that the forward points are quoted as a bid/ask pair, reflecting the appropriate interest differential applied to sellers and buyers and the additional cost (spread) quoted by the foreign exchange forwards market maker to compensate them for their quoting risk. The tomnext outrights are computed as
\[
\begin{split}
 bid_t^{tn} & = bid_t^{spot} + ask_t^{fpts} \phi \\
 ask_t^{tn} & = ask_t^{spot} + bid_t^{fpts} \phi.
\end{split}
\]
As an example of rolling a long GBPUSD position forward, the tomnext swap would involve selling GBPUSD at $bid_t^{spot}$ and repurchasing it at $ask_t^{tn}$. The cost of this \textit{roll} is thus $notional \times (bid_t^{spot} - ask_t^{tn})$, where $notional$ denotes the size of the position taken by the trader. If a trader is short GBPUSD, then to roll the position forward, she would buy $ask_t^{spot}$ and sell forward $bid_t^{tn}$, with the funding cost being $notional \times (bid_t^{tn} - ask_t^{spot})$. This funding may be a loss but also a profit. In addition, many currency market participants hold foreign exchange deliberately to capture the favourable interest rate differential between two currency pairs. This approach is known as the carry trade and is extremely popular with the retail public in Japan, where the Yen interest rates have been historically low relative to other countries for quite some time.

\Figure[!t]()[width=0.9\columnwidth]{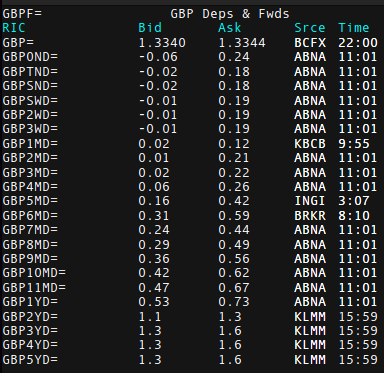}
 {Refinitiv GBPUSD forward rates.\label{fig:gbpusd_fwd_rates}}

\section{Experiment Methods} \label{sec:methods}
This section describes how our recurrent reinforcement learner targets a position directly. In addition, we also describe the baseline models that are used for comparison and contrast. Next, we explore online inductive transfer learning, with feature representation transfer from a radial basis function network to a direct, recurrent reinforcement learning agent. The radial basis function network consists of hidden processing units of the Gaussian mixture model. The recurrent reinforcement learning agent learns the desired risk position via the policy gradient paradigm. Finally, the agent is put to work trading the major spot market currency pairs.

\subsection{Targeting A Position With Direct Recurrent Reinforcement} \label{sec:drl}
Sharpe \cite{Sharpe2007} discusses asset allocation as a function of expected utility maximisation, where the utility function may be more complex than that associated with mean-variance analysis. Denote the expected utility at time $t$ for a single portfolio constituent as
\begin{equation} \label{eq:utility}
 \upsilon_t = \mu_t - \frac{\lambda}{2}\sigma_t^2,
\end{equation}
where the expected return $\mu_t = \mathbb{E}[r_t]$ and variance of returns $\sigma_t = \mathbb{E}[r_t^2] - \mathbb{E}[r_t]^2$ may be estimated in an online fashion with exponential decay, where as before $\tau$ is an exponential decay constant
\begin{equation}\label{eq:moving_mean}
 \mu_t = \tau \mu_{t-1} + (1 - \tau) r_t,
\end{equation}
\begin{equation}\label{eq:moving_variance}
\sigma_t^2 = \tau \sigma_{t-1}^2 + (1 - \tau) (r_t - \mu_t)^2.
\end{equation}
The risk appetite constant $\lambda > 0$ can be set as a function of an investor's desired risk-adjusted return, as demonstrated by Grinold and Kahn \cite{grinold2019advances}. The information ratio is a risk-adjusted differential reward measure, where the difference is taken between the model or strategy being evaluated and a baseline or benchmark strategy with expected returns $b_t = \mathbb{E} \big[ r_t^{(b)} \big]$:
\begin{equation} \label{eq:ann_ir}
 ir_t = 252^{0.5} \times \frac{\mu_t - b_t}{\sigma_t}.
\end{equation}
The similarity to the Sharpe ratio is apparent. Setting $b_t = 0$ and substituting the non-annualised information ratio into the quadratic utility and differentiating with respect to the risk, we obtain a suitable value for the risk appetite parameter:
\begin{equation}\label{eq:optimal_gamma}
 \begin{split}
 ir_t & = \frac{\mu_t}{\sigma_t} \\
 \upsilon_t & = ir_t \times \sigma_t - \frac{\lambda}{2}\sigma_t^2 \\
 \frac{d\upsilon_t}{d\sigma_t} & = ir_t - \lambda \sigma_t = 0 \\
 \lambda & = \frac{ir_t}{\sigma_t}.
 \end{split}
\end{equation}
The net returns whose expectation and variance we seek to learn are decomposed as
\begin{equation} \label{eq:drl_returns}
 r_t = \Delta p_t f_{t-1} - \delta_t |\Delta f_t| + \kappa_t f_t, 
\end{equation}

where $\Delta p_t$ is the change in reference price, typically a mid-price
\[
\Delta p_t = 0.5 \times (bid_t + ask_t - bid_{t-1} - ask_{t-1}),
\]
$\delta_t$ represents the execution cost for a price taker
\[
\delta_t = max[0.5 \times (ask_t - bid_t), 0],
\]
$\kappa_t$ is the profit or loss of rolling the overnight foreign exchange position, the so-called 'carry' (see section \ref{sec:fx_trading}) and $f_t$ is the desired position learnt by the recurrent reinforcement learner
\begin{equation}\label{eq:drl_pos}
 f_t = tanh \big ( \boldsymbol{\theta}_t^T\mathbf{x}_t \big ).
\end{equation}
The model is maximally short when $f_t = -1$ and maximally long when $f_t = 1$. The recurrent nature of the model occurs in the input feature space where the previous position is fed to the model input
\begin{equation} \label{eq:recurrent_hpu}
 \mathbf{x}_t = [1, \phi_1(\mathbf{u}_t), ..., \phi_m(\mathbf{u}_t), f_{t-1}]^T \in \mathbb{R}^{m+2},
\end{equation}
and $\phi_j(.)$ denotes a radial basis function hidden processing unit, in a network of $m$ such units, which takes as input a feature vector $\mathbf{u}_t$, see section \ref{sec:rbf}. The goal of our recurrent reinforcement learner is to maximise the utility in equation \ref{eq:utility} by targeting a position in equation \ref{eq:drl_pos}. To do this, one may apply an online stochastic gradient ascent update
\[
\boldsymbol{\theta}_t = \boldsymbol{\theta}_{t-1} + \eta \nabla \upsilon_t \equiv \Delta \boldsymbol{\theta}_t + \eta \frac{d\upsilon_t}{d\boldsymbol{\theta}_t}.
\]
Instead of a static learning rate $\eta$, one may consider the Adam optimiser of Kingma and Ba \cite{kingma2017adam}, where an adaptive learning rate is applied. This adaptive learning rate is a function of the gradient expectation and variance. The weight update then takes the form
\[
\mathbf{m}_t = \beta_1 \mathbf{m}_{t-1} + (1 - \beta_1) \nabla \upsilon_t
\]
\[
\mathbf{v}_t = \beta_2 \mathbf{v}_{t-1} + (1 - \beta_2) (\nabla \upsilon_t)^2
\]
\[
\boldsymbol{\theta}_t = \boldsymbol{\theta}_{t-1} + \eta \frac{\hat{\mathbf{m}}_t}{\hat{\mathbf{v}}_t^{0.5} + \epsilon},
\]
with $\hat{\mathbf{m}}_t = \mathbf{m}_t / (1-\beta_1)$ and $\hat{\mathbf{v}}_t = \mathbf{v}_t / (1 - \beta_2)$ denoting bias-corrected versions of the expected gradient and gradient variance, respectively. $\beta_1$ and $\beta_2$ are exponential decay constants. In earlier work, Bottou \cite{bottou-2010} had considered approximating the Hessian of the performance measure with respect to the model weights as a function of gradient only information. In practice, we find that Adam takes many iterations of model fitting to get the weights large enough to take a meaningful position via function \ref{eq:drl_pos}; this is not necessarily an Adam problem, but a result of the $\tanh$ position function taking a while to saturate. If the weights are too small, then the average position taken by the recurrent reinforcement learner will be small as well. Therefore, we settle on an extended Kalman filter \cite{Williams_rnn_ekf,Haykin2001} gradient-based weight update, albeit modified for reinforcement learning in this context.

\SetKwInput{KwRequire}{Require}
\SetKwInput{KwInit}{Initialise} 
\SetKwInput{KwInput}{Input} 
\SetKwInput{KwOutput}{Output} 

\begin{algorithm}
\caption{extended Kalman filter}\label{algo:ekf}
\SetAlgoLined
\KwRequire{$\alpha$, $\tau$
\newline \smaller \tcp{$\alpha \geq 0$ is a Ridge penalty.}
\tcp{$0 \ll \tau \leq 1$ is an exponential decay factor.}}
\KwInit{$\boldsymbol{\theta}= \mathbf{0}_d$, $\mathbf{P} = \mathbf{I}_d / \alpha$
\newline \smaller \tcp{$\mathbf{0}_d$ is a zero vector in $\mathbb{R}^{d}$.}
\tcp{$\mathbf{P}$ is the precision matrix in $\mathbb{R}^{d \times d}$.}}
\KwInput{$\nabla \upsilon_t$}
\KwOutput{$\boldsymbol{\theta}_t$}

$z = 1 + \nabla \upsilon_t^T \mathbf{P}_{t-1} \nabla \upsilon_t / \tau$

$\mathbf{k} = \mathbf{P}_{t-1} \nabla \upsilon_t / (z \tau)$

$\boldsymbol{\theta}_t = \boldsymbol{\theta}_{t-1} + \mathbf{k}$

$\mathbf{P}_t = \mathbf{P}_{t-1} / \tau - \mathbf{k} \mathbf{k}^T z$

\end{algorithm}

In algorithm \ref{algo:ekf}, $\mathbf{P}_t$ is an approximation to $[\nabla^2 \upsilon_t]^{-1}$, the inverse Hessian of the utility function $\upsilon_t$ with respect to the model weights $\boldsymbol{\theta}_t$.

We decompose the gradient of the utility function with respect to the recurrent reinforcement learner's parameters as follows:
\begin{equation} \label{eq:drl_factored_derivs}
\begin{split}
\nabla \upsilon_t & = \frac{d\upsilon_t}{dr_t} \bigg\{ \frac{dr_t}{df_t} \frac{df_t}{d\boldsymbol{\theta}_t} + \frac{dr_t}{df_{t-1}} \frac{df_{t-1}}{d\boldsymbol{\theta}_{t-1}} \bigg\} \\
 & = \frac{d\upsilon_t}{dr_t} \Bigg\{ \frac{dr_t}{df_t} \bigg\{ \frac{\partial f_t}{\partial \boldsymbol{\theta}_t} + \frac{\partial f_t}{\partial f_{t-1}} \frac{\partial f_{t-1}}{\partial \boldsymbol{\theta}_{t-1}} \bigg\} \\
 & + \frac{dr_t}{df_{t-1}} \bigg\{ \frac{\partial f_{t-1}}{\partial \boldsymbol{\theta}_{t-1}} + \frac{\partial f_{t-1}}{\partial f_{t-2}} \frac{\partial f_{t-2}}{\partial \boldsymbol{\theta}_{t-2}} \bigg\} \Bigg\}.
\end{split}
\end{equation}
The constituent derivatives for the left half of equation \ref{eq:drl_factored_derivs} are:
\[
\begin{split}
 \frac{d\upsilon_t}{dr_t} & = (1-\eta)[1 - \lambda(r_t - \mu_t)] \\
 \frac{dr_t}{df_t} & = -\delta_t \times sign(\Delta f_t) + \kappa_t \times sign(f_t) \\
 \frac{df_t}{d\boldsymbol{\theta}_t} & = \mathbf{x}_t [1 - \tanh^2{(\boldsymbol{\theta}_t^T \mathbf{x}_t)}] \\
 & + \boldsymbol{\theta}_{t,m+2}[1 - \tanh^2{(\boldsymbol{\theta}_t^T \mathbf{x}_t)}] \\ 
 & \times \mathbf{x}_{t-1} [1 - \tanh^2{(\boldsymbol{\theta}_{t-1}^T \mathbf{x}_{t-1})}].
\end{split}
\]

\subsection{Radial Basis Function Networks} \label{sec:rbf}
In Borrageiro et al. \cite{borrageiro2021online}, the authors show that online transfer learning via radial basis function networks provides a residual benefit in forecasting non-stationary time series. The residual benefit stems from the feature representation transfer of clustering algorithms. These algorithms are adapted sequentially, as are the supervised learners, which map the clustered feature space to the targets. The feature engineering that we use in this paper uses clusters formed of Gaussian mixture models. The network size is determined by the unsupervised learning procedure of finite mixture models described by Figueiredo and Jain \citep{FigueiredoJain}. Finally, we briefly describe the key ingredients of this meta-algorithm here.

The radial basis function network is a network of $m > 0$ Gaussian basis functions
\[ 
\phi_j(\mathbf{u}) = \exp{\left( -\frac{1}{2}(\mathbf{u} - \boldsymbol{\mu}_j)^T \boldsymbol{\Sigma}_j^{-1} (\mathbf{u} - \boldsymbol{\mu}_j) \right)}.
\]
Here we learn the $j'th$ mean $\boldsymbol{\mu}_j$ and covariance $\boldsymbol{\Sigma}_j$ through a Gaussian mixture model fitting procedure. Denote the probability density function of a $k$ component mixture as
\[
p(\mathbf{u}|\boldsymbol{\theta}) = \sum_{j=1}^{k}\pi_j p(\mathbf{u}|\boldsymbol{\theta}_j) = \sum_{j=1}^{k}\pi_j \mathcal{N}(\mathbf{u}|\mathbf{\mu_j}, \mathbf{\Sigma}_j),
\]
where 
\begin{equation} \label{mvn}
 \mathcal{N}(\mathbf{u}|\boldsymbol{\mu}, \mathbf{\Sigma}) = \frac{1}{(2\pi)^{d/2} |\mathbf{\Sigma}|^{1/2}}\exp \bigg[ -\frac{1}{2}(\mathbf{u} - \boldsymbol{\mu})^T \mathbf{\Sigma}|^{-1} (\mathbf{u} - \boldsymbol{\mu}) \bigg], 
\end{equation}
and the mixing weights satisfy $0 \leq \pi_j \leq 1$, $\sum_{j=1}^{k}\pi_j = 1$. The maximum likelihood estimate
\[
\boldsymbol{\theta}_{ML} = \arg \max_{\boldsymbol{\theta}} \ln p(\mathbf{u}| \boldsymbol{\theta}),
\]
and the Bayesian maximum a posteriori criterion
\[
\boldsymbol{\theta}_{MAP} = \arg \max_{\boldsymbol{\theta}} \ln p(\mathbf{u}| \boldsymbol{\theta}) + \ln p(\boldsymbol{\theta}),
\]
cannot be found analytically. The standard way of estimating $\boldsymbol{\theta}_{ML}$ or $\boldsymbol{\theta}_{MAP}$ is the expectation-maximisation algorithm \cite{Dempster1977}. This iterative procedure is based on the interpretation of $\mathbf{u}$ as incomplete data. The missing part for finite mixtures is the set of labels $\mathcal{Z} = {\mathbf{z}_0, ..., \mathbf{z}_n}$, which accompany the training data $\mathbf{u}_0, ..., \mathbf{u}_n$, indicating which component produced each training vector. Following Murphy \cite{MurphyKevinP2012Mlap}, let us define the complete data log-likelihood to be
\[
\ell_c(\boldsymbol{\theta}) = \sum_{i=1}^{n}\ln{p(\mathbf{u}_i, \mathbf{z}_i| \boldsymbol{\theta})},
\]
which cannot be computed since $\mathbf{z}_i$ is unknown. Thus, let us define an auxiliary function
\[
\mathcal{Q}(\boldsymbol{\theta}, \boldsymbol{\theta}_{t-1}) = \mathbb{E}[\ell_c(\boldsymbol{\theta})|\mathbf{u}, \boldsymbol{\theta}_{t-1}],
\]
where $t$ is the current time step. The expectation is taken with respect to the old parameters $\boldsymbol{\theta}_{t-1}$ and the observed data $\mathbf{u}$. Denote as $r_{ic} = p(z_i=c|\mathbf{u}_i, \boldsymbol{\theta}_{t-1})$, cluster $c$'s responsibility for datum $i$. The expectation step has the following form
\[
r_{ic} = \frac{\pi_c p(\mathbf{u}_i|\boldsymbol{\theta}_{c, t-1})}{\sum_{j=1}^{k} \pi_{j} p(\mathbf{u}_i|\boldsymbol{\theta}_{j, t-1})}.
\]
The maximisation step optimises the auxiliary function $\mathcal{Q}$ with respect to $\boldsymbol{\theta}$
\[
\boldsymbol{\theta}_t = \arg \max_{\boldsymbol{\theta}} \mathcal{Q}(\boldsymbol{\theta}, \boldsymbol{\theta}_{t-1}). 
\]
The $c'th$ mixing weight is estimated as
\[
\pi_c = \frac{1}{n}\sum_{i=1}^{n}r_{ic} = \frac{r_c}{n}.
\]
The parameter set $\boldsymbol{\theta}_c = \{ \boldsymbol{\mu}_c, \boldsymbol{\Sigma}_c \}$ is then
\[
\begin{split}
 \boldsymbol{\mu}_c & = \frac{\sum_{i=1}^{n} r_{ic} \mathbf{u}_i }{r_c} \\
 \boldsymbol{\Sigma}_c & = \frac{\sum_{i=1}^{n} r_{ic} (\mathbf{u}_i - \boldsymbol{\mu}_c)(\mathbf{u}_i - \boldsymbol{\mu}_c)^T }{r_c}.
\end{split}
\]
As discussed by Figueiredo and Jain \cite{FigueiredoJain}, expectation-maximisation is highly dependent on initialisation. They highlight several strategies to ameliorate this problem, such as multiple random starts, final selection based on the maximum likelihood of the mixture, or k-means based initialisation. However, the distinction between model-class selection and model estimation in mixture models is unclear. For example, a 3 component mixture in which one of the mixing probabilities is zero is indistinguishable for a 2 component mixture. They propose an unsupervised algorithm for learning a finite mixture model from multivariate data. Their approach is based on the philosophy of minimum message length encoding \cite{WallaceDowe}, where one aims to build a short-code that facilitates a good data generation model. Their algorithm can select the number of components and, unlike the standard expectation-maximisation algorithm, does not require careful initialisation. The proposed method also avoids another drawback of expectation-maximisation for mixture fitting: the possibility of convergence toward a singular estimate at the boundary of the parameter space. Denote the optimal mixture parameter set
\[
\boldsymbol{\theta}^* = \arg \min_{\boldsymbol{\theta}} \ell_{FJ}(\boldsymbol{\theta}, \mathbf{u}),
\]
where
\[
\begin{split}
 \ell_{FJ}(\boldsymbol{\theta}, \mathbf{u}) & = \frac{n}{2}\sum_{j=1}^{k} \ln \bigg( \frac{n \pi_k}{12} \bigg) + \frac{k}{2} \ln \bigg( \frac{n}{12} \bigg) \\
 & + \frac{k(n+1)}{2} - \ln{p(\mathbf{u}|\boldsymbol{\theta})}.
\end{split}
\]
This leads to a modified maximisation step
\begin{align*}
\pi_c = \frac{max \bigg\{ 0, \big( \sum_{i=1}^{n} r_{ic} \big) -\frac{n}{2} \bigg\}}{\sum_{j=1}^{k} max \bigg\{ 0, \big( \sum_{i=1}^{n} r_{ij} \big) -\frac{n}{2} \bigg\}} \\
for\ c = 1, 2, ..., k.
\end{align*}
The maximisation step is identical to expectation-maximisation, except that the $c'th$ parameter set $\boldsymbol{\theta}_c$ is only estimated when $\pi_c > 0$ and $\boldsymbol{\theta}_c$ is discarded from $\boldsymbol{\theta}^*$ when $\pi_c = 0$. A distinctive feature of the modified maximisation step is that it leads to component annihilation; this prevents the algorithm from approaching the boundary of the parameter space. In other words, if one of the mixtures is not supported by the data, it is annihilated.

We finish the section by showing figure \ref{fig:rbf_net}, which provides a visual representation of the feature representation transfer from the radial basis function network to the recurrent reinforcement learning agent. The external input to the transfer learner, represented by the left-most black circles, is a vector of daily returns of the 36 currency pairs used in the experiment, detailed in section \ref{sec:data}. The grey circles represent the radial basis function network hidden processing unit layer. In addition, we have a blue circle that represents the previously estimated position of the recurrent reinforcement learning agent. The outputs of this hidden layer are stored in a feature vector, as shown by equation \ref{eq:recurrent_hpu}. These outputs are fed into the recurrent reinforcement learning agent, who learns the position function using equation \ref{eq:drl_pos}. The weights of the position function are fitted via the extended Kalman filter procedure of algorithm \ref{algo:ekf}. The gradient vector, fed into the extended Kalman filter, is computed using equation \ref{eq:drl_factored_derivs}. This output is fed back into the hidden layer in a recurrent manner represented by the dotted blue line.

\Figure[!t]()[width=0.99\columnwidth]{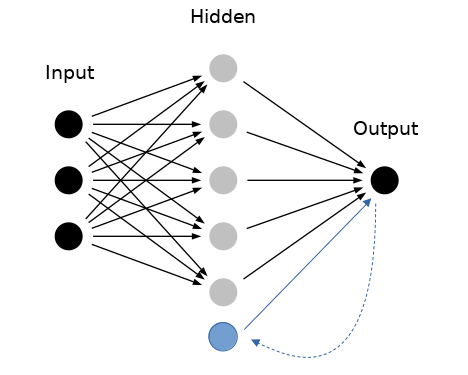}
 {Feature representation transfer from a radial basis function network to a recurrent reinforcement learning agent.\label{fig:rbf_net}}

\subsection{Baseline Models} \label{sec:baselines}
In order to assess the comparative strength of the model of section \ref{sec:drl}, we employ two baseline models. The first model is a momentum trader, which uses the sign of the next step ahead return forecast as a target position. This model is also a radial basis function network; except here, the feature representation transfer of the Gaussian mixture model cluster is made available to an exponentially weighted recursive least-squares supervised learner. A visual representation of the model is similar to figure \ref{fig:rbf_net}, without any recurrent position unit as represented by the blue circle.

\SetKwInput{KwRequire}{Require}
\SetKwInput{KwInit}{Initialise} 
\SetKwInput{KwInput}{Input} 
\SetKwInput{KwOutput}{Output} 

\begin{algorithm}
\caption{exponentially weighted recursive least-squares} \label{algo:ewls}
\SetAlgoLined
\KwRequire{$\alpha$, $\tau$
\newline \smaller \tcp{$\alpha \geq 0$ is a Ridge penalty.}
\tcp{$0 \ll \tau \leq 1$ is an exponential decay factor.}}
\KwInit{$\mathbf{w} = \mathbf{0}_d$, $\mathbf{P} = \mathbf{I}_d / \alpha$
\newline \smaller \tcp{$\mathbf{0}_d$ is a zero vector in $\mathbb{R}^{d}$.}
\tcp{$\mathbf{P}$ is the precision matrix in $\mathbb{R}^{d \times d}$.}}
\KwInput{$\mathbf{x}_{t-1}, \mathbf{x}_t \in \mathbb{R}^{d}$, $y_t$
\newline \smaller \tcp{$y_t$ is the daily sampled return of the target.}}
\KwOutput{$\hat{y}_t$}

$r = 1 + \mathbf{x}_{t-1}^T \mathbf{P}_{t-1} \mathbf{x}_{t-1} / \tau$

$\mathbf{k} = \mathbf{P}_{t-1} \mathbf{x}_{t-1} / (r \tau)$

$\mathbf{w}_t = \mathbf{w}_{t-1} + \mathbf{k} (y_t - \mathbf{w}_{t-1} ^T \mathbf{x}_{t-1})$

$\mathbf{P}_t = \mathbf{P}_{t-1} / \tau - \mathbf{k} \mathbf{k}^T r$

$\mathbf{P}_t = \mathbf{P}_t \tau$ \smaller \tcp{variance stabilisation}

$\hat{y}_t = \mathbf{w}_t ^ T \mathbf{x}_t$
\end{algorithm}

The exponentially weighted recursive least-squares fitting procedure is shown compactly in algorithm \ref{algo:ewls}.
The precision matrix $\mathbf{P}_0$ may be initialised to the identity matrix scaled by the inverse of the Ridge penalty, $\mathbf{I}_d \alpha^{-1}$ and the initial weights $\mathbf{w}_0$ are typically initialised to the zero vector. The discount factor $\tau$ is typically close to but less than 1. The particular model form is experimented with by Borrageiro et al. \cite{borrageiro2021online} in a multi-step horizon forecasting context.

Our second baseline is the carry trader, hoping to earn the positive differential overnight foreign exchange rate. Denoting the long and short carry as
\[
\begin{split}
 \kappa_t^{long} & = bid_t^{spot} - ask_t^{tn} \\
 \kappa_t^{short} & = bid_t^{tn} - ask_t^{spot} ,
\end{split}
\]
where the superscript \textit{spot} denotes the cash price, and the superscript \textit{tn} denotes the tomorrow/next price, the position of the carry trader is
\[
 f_t^{carry}= 
\begin{cases}
 sign \big( \kappa_t^{long} - \kappa_t^{short} \big) ,& \text{if } \kappa_t^{long} \text{or } \kappa_t^{short} > 0\\
 0 & \text{otherwise.}
\end{cases}
\]
In other words, the carry trader goes long the base currency if the base currency has an overnight interest rate higher than the counter currency. Equally, the carry trader sells the base currency short if the base currency has an overnight interest rate that is lower than the counter currency. Long and short carry may be a cost rather than a profit due to the bid/ask spread that traders make markets in tomnext swaps. Therefore we allow the carry trader to abstain from trading completely in such circumstances.

\section{Experiment Design} \label{sec:experiment_design}
In this section, we establish the design of the experiment, beginning with a description of the data we use and finishing up with a description of the performance evaluation criteria.

\subsection{The Data} \label{sec:data}
We obtain our experiment data from Refinitiv. We extract daily sampled data for 36 major cash foreign exchange pairs with available tomnext forward points and outrights. These foreign exchange pairs are listed in table \ref{tab:fx_pairs_static_data}. Summary statistics of the distribution of the daily returns for these currency pairs are shown in table \ref{tab:fx_pairs_desc_stats}. The dataset begins 2010-12-07 and ends on 2021-10-21, a total of 2,840 observations per pair. Daily spot mid-price returns are constructed for each of these currency pairs. These are used as the features for the recurrent reinforcement learning agent and the exponentially weighted recursive least-squares momentum trader. The mid-price is defined in equation \ref{eq:midprice}, and the return for the $k'th$ pair is simply
\[
ret_t^k = \frac{mid_t^{k}}{mid_{t-1}^{k}} - 1 \approx \ln \big( mid_t^{k}/mid_{t-1}^{k} \big).
\]

\begin{table}
\caption{The major foreign exchange pairs we use in our experiment, with Refinitiv information codes (\textit{ric})s.}
\centering
\begin{tabular}{rrrr}
\toprule
ISO currency pair & ric & tn ric \\
\midrule
AUDUSD & AUD= & AUDTN= \\
EURAUD & EURAUD= & EURAUDTN= \\
EURCHF & EURCHF= & EURCHFTN= \\
EURCZK & EURCZK= & EURCZKTN= \\
EURDKK & EURDKK= & EURDKKTN= \\
EURGBP & EURGBP= & EURGBPTN= \\
EURHUF & EURHUF= & EURHUFTN= \\
EURJPY & EURJPY= & EURJPYTN= \\
EURNOK & EURNOK= & EURNOKTN= \\
EURPLN & EURPLN= & EURPLNTN= \\
EURSEK & EURSEK= & EURSEKTN= \\
EURUSD & EUR= & EURTN= \\
GBPUSD & GBP= & GBPTN= \\
NZDUSD & NZD= & NZDTN= \\
USDCAD & CAD= & CADTN= \\
USDCHF & CHF= & CHFTN= \\
USDCNH & CNH= & CNHTN= \\
USDCZK & CZK= & CZKTN= \\
USDDKK & DKK= & DKKTN= \\
USDHKD & HKD= & HKDTN= \\
USDHUF & HUF= & HUFTN= \\
USDIDR & IDR= & IDRTN= \\
USDILS & ILS= & ILSTN= \\
USDINR & INR= & INRTN= \\
USDJPY & JPY= & JPYTN= \\
USDKRW & KRW= & KRWTN= \\
USDMXN & MXN= & MXNTN= \\
USDNOK & NOK= & NOKTN= \\
USDPLN & PLN= & PLNTN= \\
USDRUB & RUB= & RUBTN= \\
USDSEK & SEK= & SEKTN= \\
USDSGD & SGD= & SGDTN= \\
USDTHB & THB= & THBTN= \\
USDTRY & TRY= & TRYTN= \\
USDTWD & TWD= & TWDTN= \\
USDZAR & ZAR= & ZARTN= \\

\bottomrule
\end{tabular}
\label{tab:fx_pairs_static_data}
\end{table}

\begin{table}
\caption{A statistical summary of the daily returns of the major foreign exchange pairs we use in our experiment. The dataset begins 2010-12-07 and ends on 2021-10-21, a total of 2,840 observations per pair. The 25\'th, 50\'th and 75\'th percentiles of the returns distribution are shown along with the mean returns and their standard deviations.}
\centering
\begin{tabular}{lrrrrrr}
\toprule
{} & mean & std & 25\% & 50\% & 75\% \\
\midrule
USDSGD & 0.00002 & 0.00328 & -0.00178 & -0.00007 & 0.00177 \\
USDHKD & 0.00000 & 0.00034 & -0.00010 & 0.00001 & 0.00010 \\
USDIDR & 0.00017 & 0.00394 & -0.00107 & 0.00000 & 0.00150 \\
USDTHB & 0.00004 & 0.00300 & -0.00160 & 0.00000 & 0.00165 \\
USDINR & 0.00019 & 0.00441 & -0.00197 & 0.00000 & 0.00212 \\
USDKRW & 0.00003 & 0.00500 & -0.00282 & 0.00000 & 0.00277 \\
EURGBP & 0.00001 & 0.00498 & -0.00286 & 0.00000 & 0.00265 \\
USDCAD & 0.00008 & 0.00471 & -0.00272 & 0.00007 & 0.00276 \\
EURUSD & -0.00003 & 0.00511 & -0.00297 & 0.00000 & 0.00286 \\
GBPUSD & -0.00003 & 0.00546 & -0.00302 & 0.00000 & 0.00298 \\
USDJPY & 0.00012 & 0.00544 & -0.00250 & 0.00000 & 0.00290 \\
USDILS & -0.00003 & 0.00421 & -0.00244 & -0.00017 & 0.00226 \\
EURHUF & 0.00011 & 0.00449 & -0.00227 & 0.00000 & 0.00245 \\
USDZAR & 0.00031 & 0.00985 & -0.00581 & -0.00005 & 0.00599 \\
EURCZK & 0.00001 & 0.00304 & -0.00116 & -0.00004 & 0.00106 \\
AUDUSD & -0.00008 & 0.00633 & -0.00384 & 0.00002 & 0.00378 \\
NZDUSD & 0.00000 & 0.00673 & -0.00389 & -0.00007 & 0.00417 \\
USDCHF & -0.00001 & 0.00638 & -0.00293 & 0.00005 & 0.00290 \\
USDNOK & 0.00014 & 0.00721 & -0.00418 & -0.00008 & 0.00384 \\
USDSEK & 0.00010 & 0.00640 & -0.00373 & -0.00000 & 0.00367 \\
USDMXN & 0.00020 & 0.00791 & -0.00428 & -0.00004 & 0.00429 \\
USDDKK & 0.00006 & 0.00510 & -0.00286 & 0.00000 & 0.00294 \\
USDPLN & 0.00012 & 0.00730 & -0.00405 & 0.00000 & 0.00401 \\
USDHUF & 0.00017 & 0.00756 & -0.00401 & 0.00004 & 0.00426 \\
USDTRY & 0.00070 & 0.00913 & -0.00379 & 0.00029 & 0.00466 \\
USDRUB & 0.00034 & 0.01036 & -0.00428 & 0.00008 & 0.00457 \\
USDCZK & 0.00008 & 0.00643 & -0.00346 & 0.00000 & 0.00346 \\
EURSEK & 0.00004 & 0.00406 & -0.00242 & 0.00002 & 0.00234 \\
EURDKK & -0.00000 & 0.00019 & -0.00009 & 0.00000 & 0.00009 \\
EURNOK & 0.00009 & 0.00546 & -0.00280 & -0.00006 & 0.00265 \\
USDTWD & -0.00002 & 0.00280 & -0.00124 & 0.00000 & 0.00124 \\
EURJPY & 0.00008 & 0.00611 & -0.00311 & 0.00011 & 0.00330 \\
EURPLN & 0.00006 & 0.00418 & -0.00214 & -0.00002 & 0.00217 \\
EURCHF & -0.00006 & 0.00529 & -0.00147 & -0.00006 & 0.00138 \\
EURAUD & 0.00007 & 0.00575 & -0.00340 & -0.00012 & 0.00314 \\
USDCNH & -0.00001 & 0.00241 & -0.00103 & -0.00003 & 0.00099 \\
\bottomrule
\end{tabular}
\label{tab:fx_pairs_desc_stats}
\end{table}

One of the challenges that the models will face in the experiment is that these daily data show the last known top of book spot and outright prices at the end of the trading day, 5 pm EST. The bid/ask spread for these prices are at their widest statistically at this time. Therefore the execution and funding costs will be more expensive; this contrasts with a trader who can execute at a more liquid time, such as 2 pm GMT. If we try to use intraday data, say data sampled minutely, Refinitiv restricts us to 41 trading days, which is not a huge sample size. Figure \ref{fig:intraday_spreads} illustrates the challenge succinctly. It shows relative intraday bid/ask spreads
\[
spread_t^{spot} = \frac{ask_t^{spot} - bid_t^{spot}}{mid_t^{spot}},
\]
for the 36 currency pairs that we experiment with. The data are sampled minutely over two months ending mid-October 2021. The global maximum bid/ask spread occurs precisely when Refinitiv samples the daily data.

\Figure[!t]()[width=0.99\columnwidth]{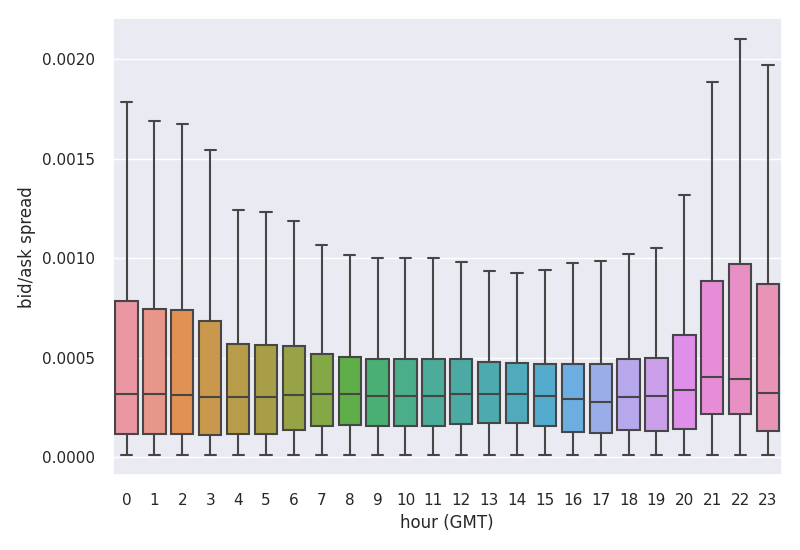}
 {Relative intra-day bid/ask spreads for the 36 Refinitiv currency pairs that we experiment with.\label{fig:intraday_spreads}}

\subsection{Performance Evaluation Methods} \label{sec:perf}
We have a little over 11 years of daily data to use in our experiment. From these data, we construct daily returns for each of the 36 currency pairs, reserving the first third as a training set and the final two-thirds as a test set. The structure of the radial basis function networks of sub-section \ref{sec:rbf} is determined in the training set, with external input being the returns of the various currency pairs. The recurrent reinforcement learning agent is also fitted in the training set to each currency pair, explicitly learning the weights in the position function \ref{eq:drl_pos}, using the extended Kalman filter learning procedure of algorithm \ref{algo:ekf}. Additionally, the momentum trader of sub-section \ref{sec:baselines} is fitted in the training set to each currency pair using algorithm \ref{algo:ewls}. Both models continue to learn online during the test set. However, the carry trader baseline does not require any model fitting.

The test set evaluates performance for each currency pair using the net profit and loss equation \ref{eq:drl_returns}. This reward, net of transaction and funding cost, is in price difference space. We convert to returns space by dividing by the mid-price computed using equation \ref{eq:midprice}. These returns are accumulated to produce the results shown in figure \ref{fig:cumul_daily_returns} and the middle sub-plots of figures \ref{fig:USDRUB_without_cost} and \ref{fig:USDRUB_with_cost}. In addition, the daily returns are described statistically in tables \ref{tab:netpnl} and \ref{tab:carrypnl}. In table \ref{tab:netpnl}, the information ratio (\textit{ir}) is computed using equation \ref{eq:ann_ir}. We set the baseline return $b_t = 0$. In summary, we evaluate performance by considering the risk-adjusted daily returns generated by each model, net of transaction and funding costs.

\subsection{Hyper-parameters} \label{sec:hyper}
The following hyper-parameters are set in the experiment:
\begin{itemize}
 \item $\tau = 0.99$; this is the exponential decay constant of moving moment equations \ref{eq:moving_mean}, \ref{eq:moving_variance}, \ref{eq:drl_returns}, extended Kalman filter weight algorithm \ref{algo:ekf} and exponentially weighted recursive least-squares algorithm \ref{algo:ewls}.
 \item $\alpha = 1$; this is the Ridge penalty of extended Kalman filter weight algorithm \ref{algo:ekf} and exponentially weighted recursive least-squares algorithm \ref{algo:ewls}.
 \item $\gamma$, the risk appetite parameter of equation \ref{eq:utility}, is initially set to $1$ and then updated by passing through the training data once and setting it via the procedure of equation \ref{eq:optimal_gamma}.
\end{itemize}

\section{Experiment Results} \label{sec:results}
Figure \ref{fig:cumul_daily_returns} shows the accumulated returns for each strategy. The reinforcement learning agent is denoted as \textit{drl}, the momentum trader is shown as \textit{mom} and the carry trader is indicated as \textit{carry}. The carry baseline performs poorly, reflecting the low-interest rate differential environment since the 2008 financial crisis. Essentially the available funding that can be earned relative to execution cost is small. Figure \ref{fig:bis_cbir} shows the direction of travel in central bank interest rates over the past 20 years. Central bank rates halved on average during the 2008 global financial crisis and have declined further since. In contrast, the momentum trader achieves the highest return with an annual compound net return of 11.7\% and an information ratio of 0.4. Additionally, the recurrent reinforcement learner achieves an annual compound net return of 9.3\%, with an information ratio of 0.52. Its information ratio is driven higher because its standard deviation of daily portfolio returns is two-thirds of the momentum trader's. Table \ref{tab:netpnl} summarises net profit and loss returns statistics by strategy, with a figure of the distribution of the daily returns in figure \ref{fig:daily_returns_hist}. Table \ref{tab:carrypnl} shows the funding or carry in returns space for each strategy. We can see that the carry baseline does indeed capture positive carry, although this return is not enough to offset the execution cost and the profit and loss associated with holding risk, which moves in a trend-following way, mainly as opposed to the funding profit and loss. How funding moves opposite to price trends is expected. Central banks invariably increase overnight rates when currencies depreciate considerably to make their currency more attractive and stem the tide of depreciation. The Turkish Lira and Russian Ruble are two cases in point. We see evidence in table \ref{tab:carrypnl} that the recurrent reinforcement learner captures more carry relative to the momentum trader. This funding capture is expected as well, as the funding profit and loss make their way into equation \ref{eq:drl_returns} and are propagated through the derivative of the utility function with respect to the model weights, using equation \ref{eq:drl_factored_derivs}.

\Figure[!t]()[width=0.99\columnwidth]{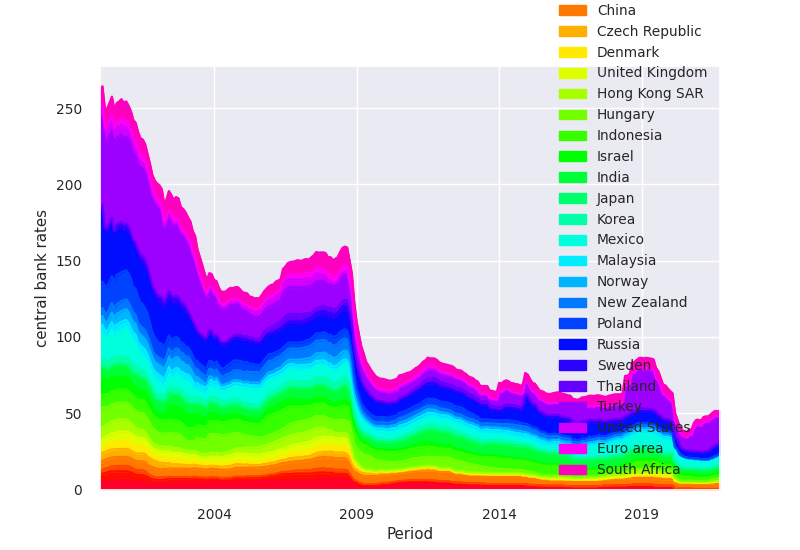}
 {Stacked central bank interest rates in percentage points, data source: Bank of International Settlements. \label{fig:bis_cbir}}

\begin{table}
\caption{Portfolio net profit and loss returns by strategy: the reinforcement learning agent is (\textbf{\textit{drl}}), momentum trader is (\textbf{\textit{mom}}) and carry trader is (\textbf{\textit{carry}}).}
\centering
\begin{tabular}{rrrr}
\toprule
{} & drl & mom & carry \\
\midrule
count & 1888 & 1888 & 1888 \\
mean & 0.00104 & 0.00121 & -0.002 \\
std & 0.032 & 0.048 & 0.052 \\
min & -0.141 & -0.202 & -0.344 \\
25\% & -0.019 & -0.028 & -0.028 \\
50\% & -0.000 & -0.002 & 0.000 \\
75\% & 0.019 & 0.028 & 0.026 \\
max & 0.245 & 0.423 & 0.200 \\
sum & 1.953 & 2.296 & -4.328 \\
ir & 0.518 & 0.403 & -0.701 \\
\bottomrule
\end{tabular}
\label{tab:netpnl}
\end{table}

\begin{table}
\caption{Portfolio funding profit and loss returns by strategy: the reinforcement learning agent is (\textbf{\textit{drl}}), momentum trader is (\textbf{\textit{mom}}) and carry trader is (\textbf{\textit{carry}}).}
\centering
\begin{tabular}{rrrr}
\toprule
{} & drl & mom & carry \\
\midrule
count & 1888 & 1888 & 1888 \\
mean & -0.00030 & -0.00050 & 0.00048 \\
std & 0.00019 & 0.00031 & 0.00036 \\
min & -0.00395 & -0.00576 & 0.00007 \\
25\% & -0.00035 & -0.00059 & 0.00029 \\
50\% & -0.00024 & -0.00040 & 0.00035 \\
75\% & -0.00019 & -0.00032 & 0.00051 \\
max & 0.00153 & 0.00072 & 0.00518 \\
sum & -0.56226 & -0.94769 & 0.90655 \\
\bottomrule
\end{tabular}
\label{tab:carrypnl}
\end{table}

\Figure[!t]()[width=0.99\columnwidth]{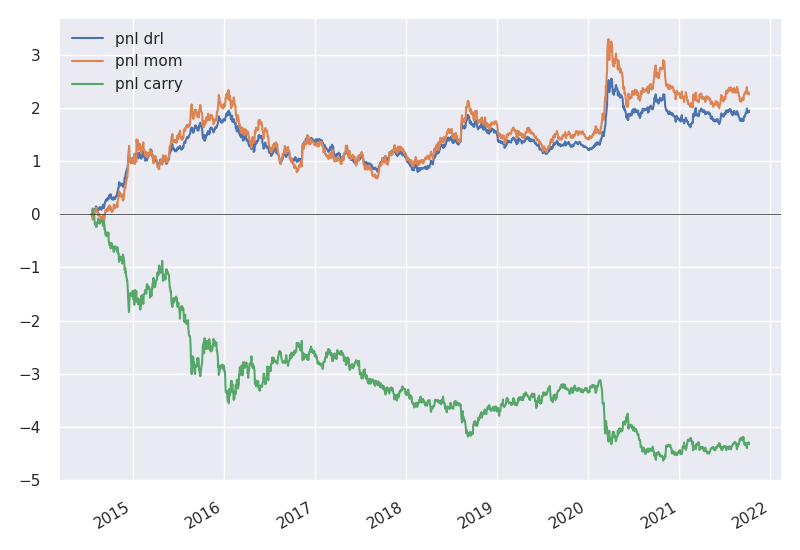}
 {Cumulative daily returns for the reinforcement learning agent (\textbf{\textit{pnl drl}}), momentum trader (\textbf{\textit{pnl mom}}) and carry trader (\textbf{\textit{pnl carry}}).\label{fig:cumul_daily_returns}}

\Figure[!t]()[width=0.99\columnwidth]{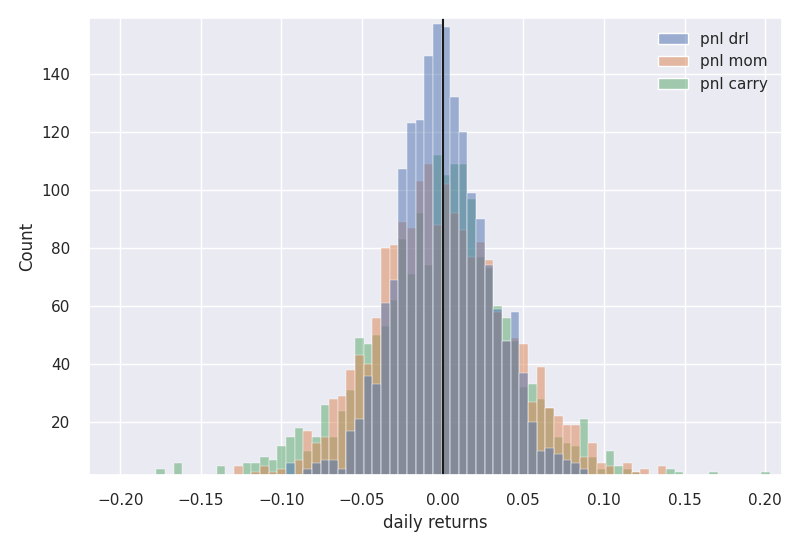}
 {Distribution of daily returns for the reinforcement learning agent (\textbf{\textit{pnl drl}}), momentum trader (\textbf{\textit{pnl mom}}) and carry trader (\textbf{\textit{pnl carry}}).\label{fig:daily_returns_hist}}

\section{Discussion} \label{sec:discussion}
Both baselines make decisions using incomplete information. The momentum trader focuses on learning the foreign exchange trends but ignores the execution and funding costs, whereas the carry trader tries to earn funding but ignores execution costs and the price movements of the underlying currency pair. In contrast, the recurrent reinforcement learner optimises the desired position as a function of market moves and funding whilst minimising execution cost. 
To demonstrate that the recurrent reinforcement learner is indeed learning from these reward inputs, we compare the realised positions of a USDRUB trader where in the former case, transaction costs and carry are removed (figure \ref{fig:USDRUB_without_cost}) and in the latter case, transaction costs and carry are included (figure \ref{fig:USDRUB_with_cost}). We see that without cost, the recurrent reinforcement learner realises a long position (buying USD and selling RUB) broadly, as the Ruble depreciates over time. In contrast, when funding cost is accurately applied, the overnight interest rate differential is roughly 6\%, and the recurrent reinforcement learner learns a short position (selling USD and buying RUB), capturing this positive carry. The positive carry is not enough to offset the rapid depreciation of the Ruble.

How significant are these results? Grinold and Kahn \cite{grinold2019advances} show table \ref{tab:empirical_ir} of empirical information ratios for US fund managers over the five years from January 2003 through December 2007. The data relates to 338 equity mutual funds, 1,679 equity long-only institutional funds, 56 equities long-short institutional funds and 537 fixed-income mutual funds. Although now a bit dated, the results indicate that our recurrent reinforcement learner that trades statistically at the worst time of day in the foreign exchange market achieves an information ratio at the $75'th$ percentile of information ratios achieved empirically by various passive and active fund managers within fixed income and equities. The momentum trader achieves an information ratio between the $50'th$ and $75'th$ percentile. The information ratio is a measure of consistency and has a probabilistic interpretation: it measures the probability that a strategy will achieve positive residual returns in every period \cite{grinold2019advances}. Equation \ref{eq:ann_ir} shows that the information ratio is the ratio of residual return to residual risk. Let us denote this residual return as the strategy's \textit{alpha}:
\[
\alpha_t = \mu_t - b_t.
\]
The probability of realising a positive residual return is
\[
Pr(\alpha_t > 0) = \Phi(ir_t),
\]
where $\Phi(.)$ denotes the cumulative normal distribution function. In this respect, we find that recurrent reinforcement learner has a probability of positive residual return of 70\% and the momentum baseline has a probability of positive residual return of 66\%.

\begin{table}
\caption{Empirical information ratios, source: Blackrock.}
\centering
\begin{tabular}{p{1cm}p{1cm}p{1cm}p{1cm}p{1cm}p{1cm}}
\toprule
asset class & equities & equities & equities & fixed income & both \\
\midrule
percentile & mutual funds & long & long short & institutional & average \\
90 & 1.04 & 0.77 & 1.17 & 0.96 & 0.99 \\
75 & 0.64 & 0.42 & 0.57 & 0.50 & 0.53 \\
50 & 0.20 & 0.02 & 0.25 & 0.01 & 0.12 \\
25 & –0.21 & –0.38 & –0.22 & –0.45 & –0.32 \\
10 & –0.62 & –0.77 & –0.58 & –0.90 & –0.72 \\
\bottomrule
\end{tabular}
\label{tab:empirical_ir}
\end{table}

In terms of future work, one might consider a multi-layer perceptron version of our recurrent reinforcement learner. One might also consider an echo state network \cite{Yildiz2012RevisitingTE} version of the model. In addition, one might be able to improve the results further by applying a portfolio overlay. The utility function of equation \ref{eq:utility} is readily treated as a portfolio problem
\[
\upsilon_t = \mathbf{h}^T\boldsymbol{\mu}_t - \frac{\lambda}{2}\mathbf{h}^T\boldsymbol{\Sigma}_t\mathbf{h},
\]
where the optimal, unconstrained portfolio weights are obtained by differentiating the portfolio utility with respect to the weight vector
\[
\mathbf{h}^* = \frac{1}{\lambda}\boldsymbol{\Sigma}_t^{-1} \boldsymbol{\mu}_t.
\]
Another approach is to treat portfolio selection as a policy gradient problem, where the policy of picking actions, or this case portfolio constituents, is estimated via function approximation techniques.

\Figure[!t]()[width=0.99\columnwidth]{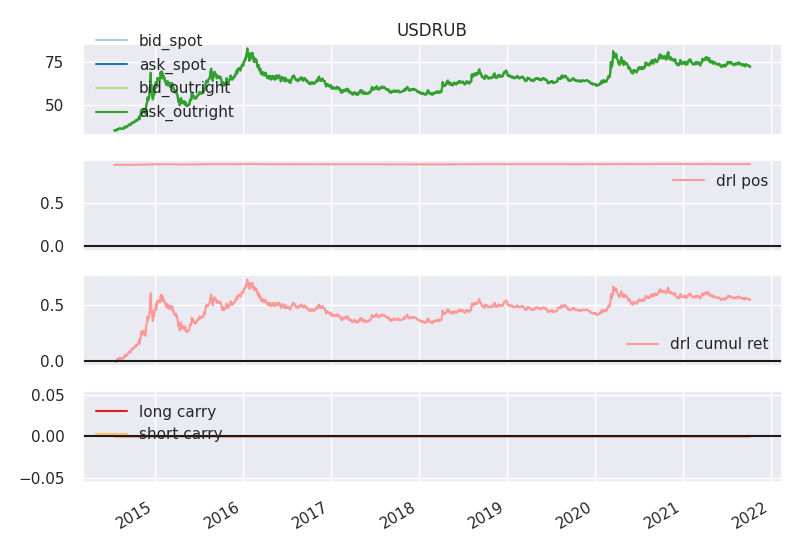}
 {A USDRUB reinforcement learning agent trading \textit{without} execution or funding cost.\label{fig:USDRUB_without_cost}}

\Figure[!t]()[width=0.99\columnwidth]{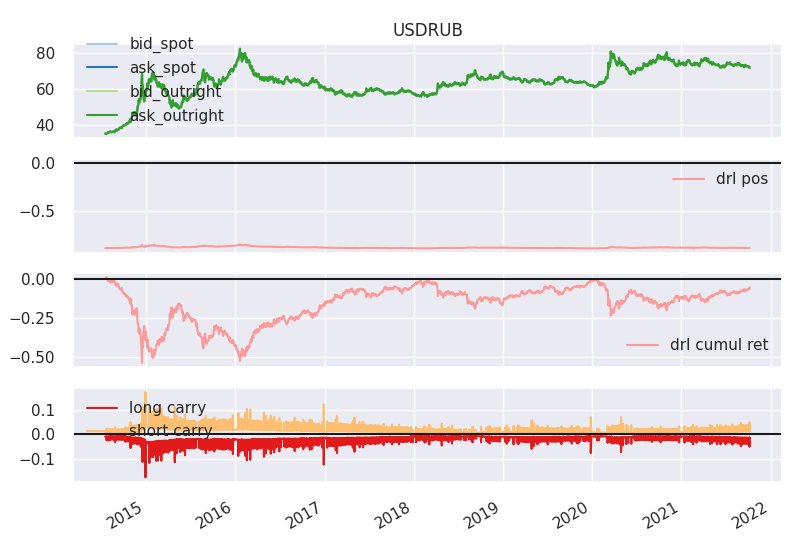}
 {A USDRUB reinforcement learning agent trading \textit{with} execution and funding cost.\label{fig:USDRUB_with_cost}}

\section{Conclusion} \label{sec:conclusion}
We conduct a detailed experiment on major cash foreign exchange pairs, accurately accounting for transaction and funding costs. These sources of profit and loss, including the price trends that occur in the currency markets, are made available to our recurrent reinforcement learner via a quadratic utility, which learns to target a position directly. We improve upon earlier work by casting the problem of learning a risk position in an online learning context. This online learning occurs sequentially in time but also via transfer learning. This transfer learning takes the form of radial basis function hidden processing units, whose means, covariances and overall size are determined by an unsupervised learning procedure for finite Gaussian mixture models. The intrinsic nature of the feature space is learnt and made available to the recurrent reinforcement learner and baseline supervised-learning momentum trader.
The recurrent reinforcement learning trader achieves an annualised portfolio information ratio of 0.52 with a compound return of 9.3\%, net of execution and funding cost, over a 7-year test set, despite forcing the model to trade at the close of the trading day 5 pm EST, when trading costs are statistically the most expensive. The momentum baseline trader achieves a similar total return but a lower risk-adjusted return. The recurrent reinforcement learner does maintain an essential advantage in that the model's weights can be adapted to reflect the different sources of profit and loss variation, including returns momentum, transaction costs and funding costs. We demonstrate this visually in figures \ref{fig:USDRUB_without_cost} and \ref{fig:USDRUB_with_cost}, where a USDRUB trading agent learns to target different positions that reflect trading in the absence or presence of cost.

\typeout{}
\EOD


\bibliographystyle{IEEEtranN} 
\bibliography{references}

\vspace{-7mm}
\begin{IEEEbiography}[{\includegraphics[width=1in,height=1.25in,clip,keepaspectratio]{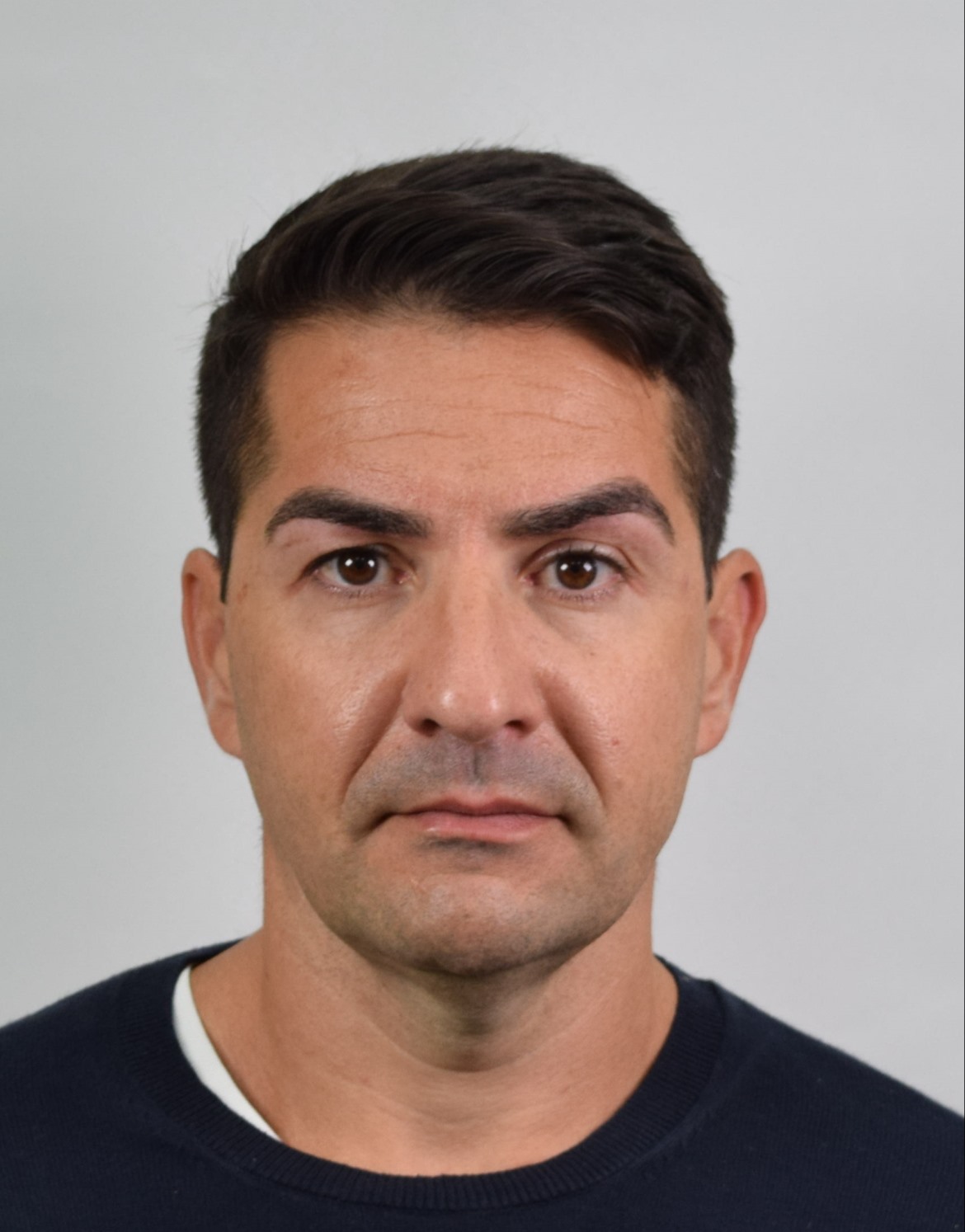}}]{Gabriel Borrageiro} is currently a PhD student at University College London, Computer Science department, in the Financial Computing Group. His research interests include online learning, reinforcement learning, transfer learning, recurrent neural networks and financial time series. Gabriel obtained his executive MBA from Cass Business School, City University, London, in 2008. He also obtained a higher national diploma from Damelin College, South Africa, in 1999. Gabriel is also employed as a quantitative researcher at BlueCrest Capital.
\end{IEEEbiography}
\vspace{-7mm}

\begin{IEEEbiography}[{\includegraphics[width=1in,height=1.25in,clip,keepaspectratio]{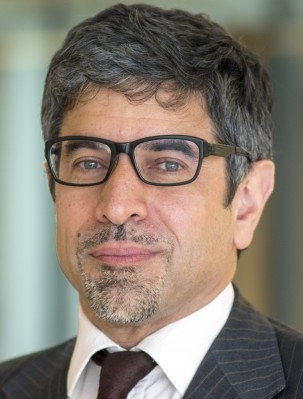}}]{Nick Firoozye} is currently Honorary Reader, Computational Finance, in the Computer Science department at University College London and also part of the Financial Computing Group. He obtained his PhD and MS in mathematics at New York University and a BS in mathematics at Harvey Mudd College. He also works for Exos Bank in the systematic rates trading business.
\end{IEEEbiography}
\vspace{-7mm}

\begin{IEEEbiography}[{\includegraphics[width=1in,height=1.25in,clip,keepaspectratio]{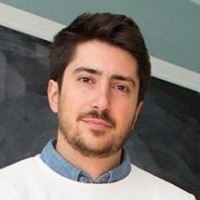}}]{Paolo Barucca} is a lecturer at University College London, Computer Science department and also part of the Financial Computing Group. He is also editor in chief of the science dissemination project, La Scienza Coatta and scientific officer of the Blockchain Education Network. Paolo received his PhD in theoretical and mathematical physics from Sapienza Universita di Roma in 2015.
\end{IEEEbiography}

\vfill

\end{document}